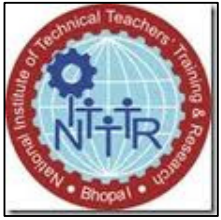

# Interactive Learning through Hands-on Practice using Electronic Mini - Lab (EML): a Case Study

**Vikas J Dongre*[1], Ramkrishna V Yenkar[1], Vijay H Mankar[1]**

[1]*Department of Electronics and Telecommunication,*
*Government Polytechnic, Nagpur, India*
*[1]Email: dongrevj@yahoo.co.in

*Abstract--* **In this paper, a new approach to impart practical skill based technical education is presented in comprehensive manner. An Electronic Mini-Lab (EML) is devised containing basic design and test instruments with electronic components, ICs, connecting wires and battery. Using the EML, students perform various formal and informal digital and analog circuit practicals as well as design prototype of projects. This gives them hands-on experience, sense of belonging and sense of cooperation. The EML is useful for performing many practicals of various subjects. The EML also reduces the workload of college laboratories. Students have their own individual EML at their disposal anytime which can be used to design hobby projects as a fun too. This will make them skilled engineers. This provides tremendous benefits in teaching learning process. It also boosted the interest, confidence of students and teachers. Incorporating active/ cooperative learning into traditional instruction can be a useful pedagogical tool to help students to perform practicals and project work any time anywhere. This concept is remarkably simple and cost effective but the dividends can be profound.**

## I. INTRODUCTION

The most desired outcome of engineering technology education is the creation of skilled technologists who are able to approach the design and application of both hardware and software with aptitude and creativity [1]. Engineering is a practical discipline. It is a hands-on profession where doing is the key. It is generally agreed that the aims of laboratory work can be broadly classified in three domains of learning [2].

• *Cognitive Domain:* Activities like Instrumentation, Modeling, Experimentation, Data Analysis and Design.

• *Psychomotor Domain:* Activities like manipulating the apparatus, Sensory Awareness.

• *Affective Domain*: behavior and attitudes learn from failure, creativity, safety, communication, teamwork, and ethics in the laboratory.

Exposing students to all three of these domains is necessary to produce an effective engineer. Cognitive research indicates that real learning and understanding is better accomplished through cooperative and interactive techniques. Furthermore, being brought up in an era of TV and video games, today's students have limited attention span but they respond well to multimedia stimuli and interactive activities [3]. Recent studies such as [4] [5] show that there is a gap between traditional training and the skills actually needed in today's job markets in the various domains like cognitive flexibility, creativity, knowledge transfer, and adaptability. Therefore, being able to solve new problems based on the knowledge acquired has become a desired outcome of technical education institutes. In order to bridge this educational gap, the long-term goal is to create a progressively more engaging laboratory experience with problem solving emphasis and various skill and knowledge acquisition. The objectives are:

• Enable active and hands-on student engagement to develop excellent problem solving and troubleshooting skills.

• Provide opportunities for the students to develop teamwork skills.

• Encourage lifelong curiosity towards science and technology.

In the context of college classroom, such active learning involves students in doing things and thinking about the things they are doing [6]. Indian National Mission on Education through Information and Communication Technology (ICT), in its objectives suggested about Facilitating, development and deployment of ultra low cost physical tool kits for engineering and science students to encourage project and design based learning complementary to the e-learning [7].

A practical in the laboratory is an essential part for verification of classroom theory in all subjects of engineering education. The ratio of theory to practical varies at various levels of education. In engineering degree, the ratio is generally 60:40. In engineering diploma, it is 40:60. In industrial training institutes (ITI) it is 20:80. This highlights the importance of practical sessions.

In this paper our work specifically emphasize on electronic engineering group branches. However, the idea can also be implemented in other branches of engineering with suitable modifications.

In electronic engineering laboratories, various instruments and components are very small and delicate. They need careful handling. Despite this, the quality of practical kits imposes problems. The kits are fitted in the opaque cabinet and terminals are brought out. While handling the trainer kits, by many batches of students and occasionally providing voltages of incorrect ranges or making wrong connections, the kits start malfunctioning. Occasionally the kits do not function at all. This makes the teachers and students loose their confidence. The maintenance engineer is not always accessible. So the practicals are not properly performed. It is found that in many institutes, 40% of the performance based practicals are converted into study practical. This deprives the students from getting practical exposure. Students start getting biased that electronic practical normally do not perform well.

Even if the new setup of kits properly operates, students are not able to see or handle all the electronic components physically. Many students study a lot about various electronics components analytically, but when examined, they are not even able to identify the basic components. In order to impart effective teaching and learning, we have introduced Electronic Mini Lab (EML) concept in Government Polytechnic Nagpur since 2008-09 [8]. Replacing conventional laboratory format, 2 hours of activity-based instruction per week provides an insight of in-depth practical exposure to large number of students [9]. This revised study format encouraged the students greatly to understand by doing, not merely listening or observing. This helped students to improve understanding of various engineering concepts [10].

## II. ELECTRONICS MINI-LAB (EML)

The proposed EML consists of all the required items like battery, ICs, LEDs, connecting wires and various other tools. The detailed list is given in table I. It may be noted that students at their own cost procure the kit by personally going to electronic market.

The EML is introduced in third semester when students start learning core technology related to their branch. Initially the EML was used to perform the practicals in Digital Electronics Lab. Therefore, the EML contains the components of this laboratory. Other components can be added in due course of time as and when they study other subjects in parallel or in later semesters.

As mentioned in table-I, the total cost of individual EML is approximately Rs.1000 ($25). The cost can be brought down if purchased in bulk.

## III. PROCUREMENT OF EML

In the first or second lecture of Digital Electronics course during third semester, students are shown a standard EML box and all its components. They are given the list of items as mentioned in Table I, along with appropriate quality, quantity and price range. They are also suggested the addresses of electronic venders. They are instructed to purchase the items in a week's time, so that Laboratory work can be undertaken soon. They are also instructed to go in groups of 4 to 5 so that they may get group discount.

The transparent plastic container with convenient size is selected which can be accommodated in the college bag of students. Recommended dimension in our college is mentioned in table I. Two small visiting card size containers were used for storing small size components and ICs. The students bring the EML during practical days

### TABLE I: COMPONENTS OF EML

| SN | Item | Qty | rate | Approx Cost (Rs) |
|---|---|---|---|---|
| 1 | Digital Multi-meter | 1 | 200 | 200 |
| 2 | Soldering Iron | 1 | 150 | 150 |
| 3 | Soldering stand | 1 | 50 | 50 |
| 4 | De-soldering Braid (mesh) | 1mtr | 10 | 10 |
| 5 | soldering wire | 1 | 25 | 25 |
| 6 | Soldering flux | 1 | 15 | 15 |
| 7 | Breadboard | 1 | 75 | 75 |
| 8 | 9 V Battery with connector | 1 | 20 | 20 |
| 9 | Wire stripper | 1 | 50 | 50 |
| 10 | Nose plier (small) | 1 | 60 | 60 |
| 11 | Hook up wire various colors | 3 mtr | 03 | 10 |
| 12 | Berg strip male and female | 2 | 05 | 10 |
| 13 | Heat shrink tube 3mm diameter | 1 mtr | 05 | 05 |
| 14 | General purpose PCB | 1 | 20 | 20 |
| 15 | Plastic container 10"x8"x2" | 1 | 50 | 50 |
| 16 | Plastic container3"x2"x0.5" | 2 | 10 | 20 |
| 17 | LED various colors 5 mm | 5 | 02 | 10 |
| 18 | Resisters ¼ watt (330 ohm) | 10 | 0.5 | 05 |
| 19 | IC 7805 , voltage regulator | 1 | 15 | 15 |
| 20 | ICs: 7400, 7402, 7404, 7408, 7432, 7486, 7493, 555, LM324, LM741 | 1 each | 15 | 150 |
| 21 | LDR, Thermister, Photodiode, POT | 2 each | 25 | 50 |
| | Total Approximate Cost Rs. | | | 1000 |

358

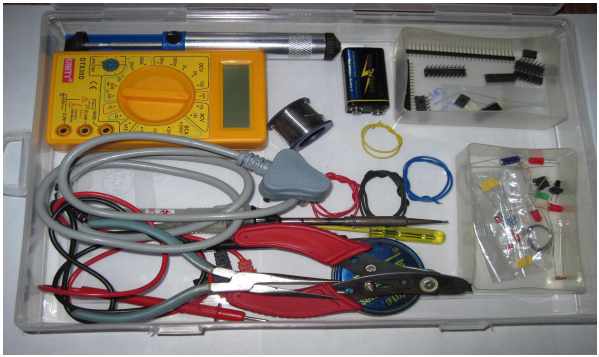

**Fig. 1: Electronic Mini-Lab (EML)**

### IV. CONNECTIONS AND WORKING WITH EML

In the first or second practical session, as and when the students come with their own EML box, they are familiarized with the arrangement available in the Breadboard. Basic workable breadboard connections for power supply are made as shown in fig. 2.

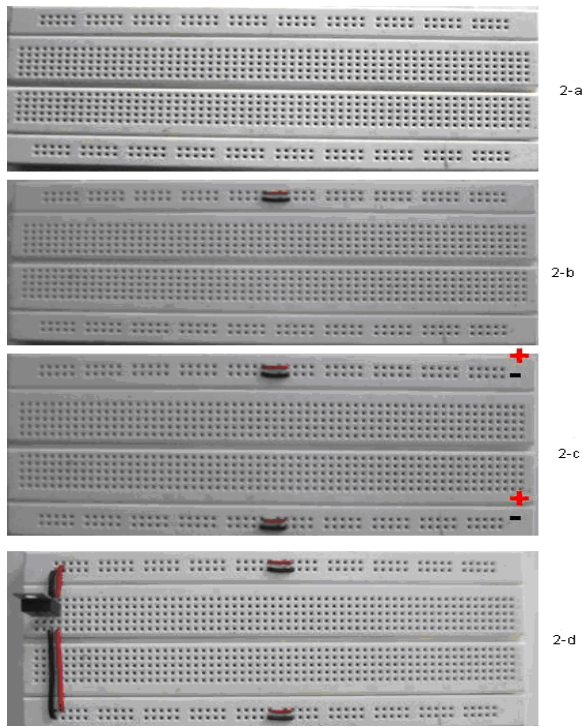

2-a: Initial Breadboard,
2-b: cascading upper horizontal lines
2-c: cascading lower horizontal lines,
2-d: Connecting power supply to power lines

**Fig 2: Basic wiring for power supply**

3 pin regulator IC7805 is connected. 9V Battery is connected at the input of regulator IC. Output indicator LED connections are made as shown in figure 3. Finally a basic logic circuit is mounted for initial testing purpose as shown in fig.4.

The students get confidence when the LED glows. The step by step procedure is instructed to all the class on blackboard and the students make the necessary connection on breadboard simultaneously. It is like hands-on workshop experience for the students. [8], [9], [11].

The connections are tightly and perfectly made with proper colour coding since these connections are to be preserved for the whole semester. The students are instructed to follow the identical connection methodology on the breadboard for the ease of testing and troubleshooting.

The horizontal short-circuited two lines on the top and bottom of the breadboard are used for $V_{CC}$ and Ground using Red and Black wires respectively. Once the LED glows, it ensures that connections, regulator IC, battery are operating properly. From next laboratory session, the actual practicals on combinational logic begin. For the entire practicals, centre line of the breadboard is used for mounting the ICs. $V_{CC}$ and ground are available on top and bottom lines both. $V_{CC}$ and

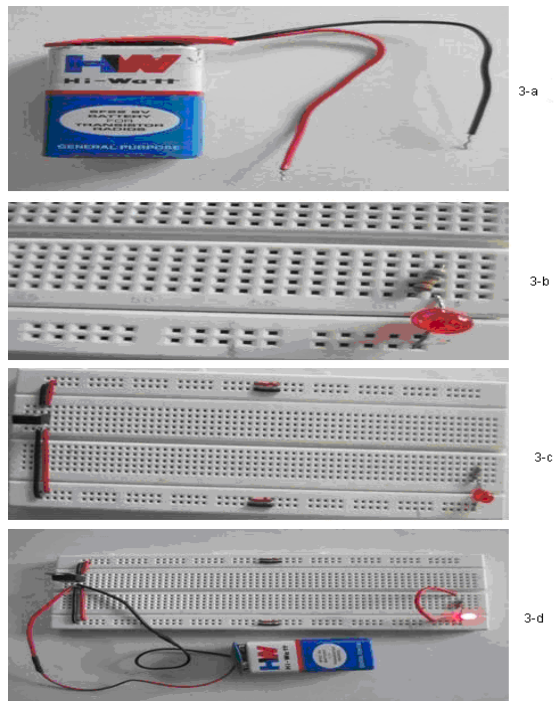

3-a: Battery with connector,
3-b: connecting LED as Output
3-c: circuit with power supply and output probe
3-d: Final basic working model

**Fig. 3 Battery Regulator and LED connection**

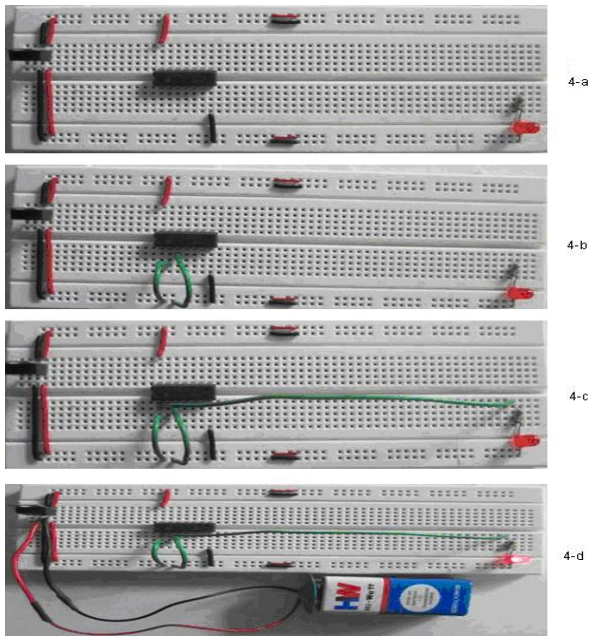

4-a: Mounting the IC,      4-b: power and signal to IC
4-c: connecting output probe to IC
4-d: final circuit with battery connection and output

**Fig. 4: Basic Logic Circuit setup for testing**

GND are suitably connected to the IC pins using hook up wires. Inputs are given from $V_{CC}$ and GND, as logic 1 and logic 0. Output of the Digital IC is given to LED through current limiting resister. Once the input and output connections are properly made, output is observed on LEDs very conveniently. This gives visible output and pleasant experience to the students. After the first successful experiment on the breadboard, students get conversant with the arrangement. From next practical onwards, the students operate the EML indigenously to design breadboard-based circuit.

The teacher discusses idea of logic circuit by drawing it on blackboard and within the work of 15 to 20 minutes, the results start coming in. The students themselves help each other to make the circuit work. Every student operates on his own circuit. Hence, there is no crowd near any specific kit.

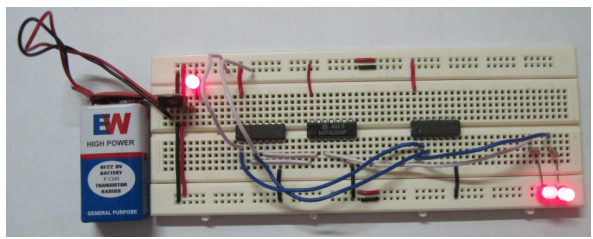

**Fig 5: Sample practical setup**

The practicals are performed only using low power battery. Since no mains power (230 V ac) supply is needed, there is no possibility of electric shock. The students and teacher gets enough time for technical discussion, clearing the doubts and continuous assessment.

For sequential circuit design, IC555 timer is connected on the breadboard in mono stable mode with 0.5-second time period. It is triggered using push to on switch for manual clock generation. Thus, all the sequential digital circuits can also be designed on the breadboard.

Wire stripper is needed occasionally for preparing the connecting wires. Berg strip (male) may be connected to both the ends of connecting wires for comfortable connections on the breadboard. This also necessitates soldering. Heat shrink tube is used for taping the solder joints.

## V. RESULT AND DISCUSSION

*A. Experiences with EML*

1. Students may interchange $V_{CC}$ and GND, which short circuits the battery. To avoid this, power indicator LED is used near power supply. If LED doesn't glow, power is immediately removed to minimize battery discharge. Heating of regulator IC or other ICs is also the indication of short circuit. Students are advised to touch the ICs occasionally. If some IC is hot abnormally, students are asked to recheck the connections.

2. Due to polarity change, IC or LEDs may get damaged. But it is not very expensive affair as ICs are cheap. Components are replaced. After such damage, students learn to make the connections more carefully.

3. Circuits may not work if ICs are loosely fitted or wrongly fitted. Each of such failure teaches the students one or the other concept. Thus learning takes place in all the cases whether result is obtained or not.

4. Not all the components in the EML may be required for every practical (e.g. Soldering Iron, multi meter). Such components may be kept at home to reduce the weight of the EML.

*B. Advantages of EML*

1. Students visit Electronic market, observe lot of components, and appreciate variety of electronic components and materials, which they may not observe in the laboratory.

2. Students understand various ranges and tolerances of components, their quality and cost.

3. As the students spend their own money, they cultivate the sense of belonging. They handle the instruments and components carefully. In this process, they also understand the importance of handling laboratory instruments carefully, bringing down the failure rate of laboratory instruments.

4. As the students handle the electronics components physically, they know their footprints, size, shape and weight. This helps them while drawing PCB layout artwork for their project in later semesters and handling other components.

5. The students gain confidence and skill developing positive attitude in them. They understand that electronic circuits operate nicely if systematically connected.

6. As the breadboard requires 9V battery and output LEDs, practicals can be performed virtually any where. This brings down the workload of laboratory and teacher. Students have freedom to perform optional practicals, hobby circuits anywhere anytime. .

7. The skill of connecting the components helps them in effectively performing practical of other subjects also.

8. EML has multi meter, soldering iron, plier, cutter and stripper. These tools are essential for an electronics engineer anytime. They are useful in other laboratories like Electronic Workshop where they fabricate their own projects. The EML is also useful in other laboratories like Network Theory, Linear integrated Circuits, Microcontroller Lab. It is very useful in designing prototype in Electronics workshop and projects [12].

9. Concept of hands-on practice using EML can equally be used in various related disciplines such as Electrical, Computer, Instrumentation and Mechanical engineering etc. where electronic courses are taught.

10. Component damage rate is only up to 20% and it reduces as students get more experience. They understand that damage of component is loss of their money enabling them to work with utmost care.

11. Failure of circuit teaches the students a lot of technical design principles, which they share with their colleagues. Thus maximum learning takes place informally too.

12. As the students get 100% practical exposure, they also develop interest in theory classes. This synergies the teaching learning process. Students now actively participate in theory and practical both. Hence teacher also gets indirectly motivated and imparts knowledge with more interest.

13. Teacher share his/her experience of EML with other teachers also. Hence other teachers also try to incorporate practical based on EML for their courses wherever possible. This increases the utility of EML. The assessment of performance of students is very simple and concise as many technical aspects can be assessed on the breadboard design.

14. The Breadboard setup used by students throughout the semester gives them opportunity to design various hands-on robotics projects also, which has become useful educational tools across a variety of subjects. Because of multidisciplinary nature, the study of robotics has become a valuable tool for the practical, hands-on application of concepts in various engineering and science topics [13-14]. One of such small robot prototype, using breadboard designed by a student is shown in fig. 7.

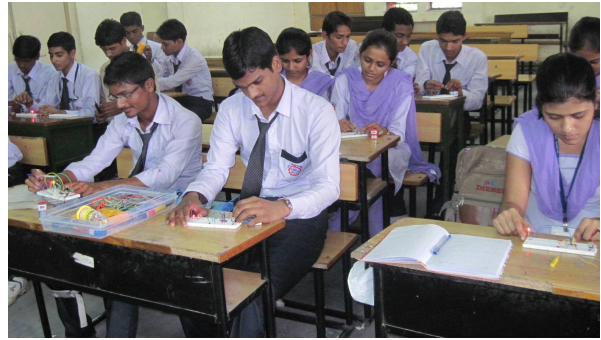

**Fig.6: Students experimenting with EML**

*C. Improvements in the students*

The EML introduced in the college enabled students with more practical exposure. The observation of students using EML clearly indicates that they had gained better practical approach for designing any circuit. The pass out students from this college, undergoing their graduation, experience that they are well equipped with the potential of practical approaches as compared to their mates coming from other colleges. This clearly shows the proposed EML proved to be very beneficial tool for students to gain hands-on practice.

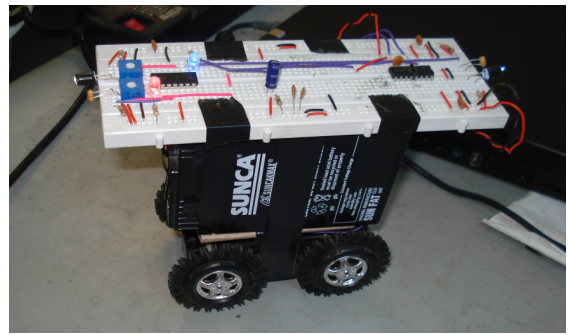

**Fig 7: Small robot using Breadboard designed by a student**

### VI. CONCLUSION

The present paper discusses the case study of Electronics Mini Lab (EML) implemented as an interactive method of teaching learning using easily available instruments and electronic components. The implemented EML is cost effective and portable. As there is equal emphasis on imparting information and developing practical skills, students accept the concept with great enthusiasm. They perform practicals with more zeal improving their psychomotor skills. Moreover, practical work of various engineering courses can be carried out using this EML

Students develop sense of belonging while handling the electronic components. They are ready to accept the new concepts in learning. Students are involved in higher order thinking (analysis, synthesis and evaluation). They are well equipped with the potential of practical approaches, which helps them in higher education and in professional career. Thus, EML is greatly useful for inculcating the skill based learning process in Engineering Diploma and undergraduate programs of Electronics Engineering curriculum. The concept can be equally extended with suitable modifications in other branches of engineering.

## AUTHORS

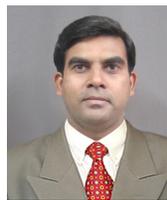

**Vikas J. Dongre** received his B.E and M.E. in Electronics in 19991 and 1994 respectively. He served as lecturer in SSVPS engineering college Dhule, (MS) India from 1992 to 1994, He Joined Government Polytechnic Nagpur in 1994 as Lecturer where he is presently working as lecturer in selection grade. His areas of interests are Microcontrollers, embedded systems, image recognition, and Education Technology. He has published two research paper in international conferences and two research papers in international journals. Presently he is pursuing for Ph.D in Offline Devnagari Character Recognition.

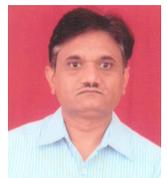

**Ramkrishna V Yenkar** received his B.E in Industrial Electronics and M. Tech. in Electronics from VNIT, Nagpur in 1987 and 1990 respectively. He served as a Lecturer about 7 years in engineering colleges. Since last 15 years, he is working as Head, Dept of Electronics, Government Polytechnic, Maharashtra State, presently at Arvi. He has published more than 10 research papers in national and international conferences. His field of interest includes Educational Technology, Analog/Digital systems and chaotic in nonlinear systems. Presently he is pursuing for Ph.D in chaotic in nonlinear systems.

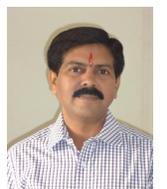

**Vijay H. Mankar** received M. Tech. degrees in Electronics Engineering from VNIT, Nagpur University, India in 1995 and Ph.D. (Engg) from Jadavpur University, Kolkata, India in 2009 respectively. He has more than 17 years of teaching experience and presently working as a Lecturer (Selection Grade) in Government Polytechnic, Nagpur (MS), India. He has published more than 40 research papers in international conference and journals. His field of interest includes digital image processing, data hiding and watermarking.